\documentclass[aps,pre,twocolumn,floatfix,nofootinbib,showpacs,amsmath,amssymb]{
revtex4}

\usepackage[english]{babel}
\usepackage{graphicx}
\usepackage{textcomp}
\usepackage{amsmath}

\begin{document}

\title{The Markov Switching Multi-fractal models as a new class of REM-like models in 1-dimensional space.}
\author{David B. Saakian$^{1,2,3}$}
\affiliation{$^1$Institute of Physics, Academia Sinica, Nankang,
Taipei 11529, Taiwan}
 \affiliation{$^2$A.I. Alikhanyan Nation al Science Laboratory (Yerevan Physics Institute) Foundation,
Alikhanian Brothers St. 2, Yerevan 375036, Armenia}
\affiliation{$^3$National Center for Theoretical Sciences:Physics
Division, National Taiwan University, Taipei 10617, Taiwan}

\date{\today}

\begin{abstract}
We  map the  Markov Switching Multi-fractal model (MSM) onto the
Random Energy Model (REM). The MSM is, like the REM, an exactly
solvable model in 1-d space with non-trivial correlation functions.
According to our results, four different statistical physics phases
are possible in random walks with multi-fractal behavior. We also
introduce the continuous branching version of the model, calculate
the moments and prove multiscaling behavior. Different phases have
different multi-scaling properties.
\end{abstract}
 \maketitle

\section{Introduction}

The Random Energy Model (REM) introduced by Derrida
\cite{de80}-\cite{de91} is one of the fundamental models of modern
physics. Originally derived as a mean-field version of spin-glass
models, it has subsequently been applied to describe some features
of 2-d  Liouville models \cite{ch96},\cite{ca97}, as well as the
properties of quenched disorder in d-dimensional space with a
logarithmic correlation function of energy disorder.

The logarithmic correlation is easy to organize both for models on a
hierarchic tree and in 2-d. The REM structure in the d-dimensional
logarithmic correlation case has been proved in \cite{ca01} using
 Bramson\rq s results \cite{br82}. In \cite{bu08},\cite{fy09} a REM-like
 model in 1-d space has been solved directly, using the
generalization of Selberg integrals \cite{fy09}. A mapping has been
used in \cite{sa09} to map the REM onto strings.

A REM can be formulated not only for the case of normal
distributions of energies, which corresponds to the logarithmic
correlation function of energies on a hierarchic tree, similar to
logarithmic correlation functions in d-dimensional disorder case
\cite{ca01}, but also for general distribution of energies
\cite{de90},\cite{sa93}. It is an open problem to find solvable
models with  the non-logarithmic correlation for energy disorder in
1-d space. To describe the fluctuations in financial market,
\cite{cf01} and \cite{cf02} have constructed some dynamical models,
the Markov switching multi-fractal models (MSM). The MSM has time
translational symmetry, contrary to cascade models, defined on
hierarchic trees \cite{ca97}.
 The connection of the 1-d REM model
\cite{fy09} with the multi-fractal random walk model
\cite{ba00}-\cite{ba06} was found in \cite{sa11}. In this article we
will prove that the dynamical models of \cite{cf01} and \cite{cf02}
provide a 1-d REM where the correlation function for energies ($\ln
u_t$ in our case) has a general character instead of being
logarithmic.

Let us give the definition of the MSM  model, following
\cite{cf01,cf02,lu08}. In the MSM model, one considers the sequence
of variables $r_t$, where $ t\ge 0$ describes a discrete moment of
time:
\begin{eqnarray}
\label{e1} r_t=x_tu_t,
\end{eqnarray}
 $x_t$ has a normal distribution,
\begin{eqnarray}
\label{e2} <x_t^2>=J^2
\end{eqnarray}
  and $u_t$ is defined at the moment t of time via a
product of k components $M(t,l)$
\begin{eqnarray}
\label{e3} u_t=\prod_{l=1}^kM(t,l).
\end{eqnarray}
The variables $M(t,l)$ are random variables with some distribution.

Every moment of time our random variables are  replaced with new
ones with a probability
\begin{eqnarray}
\label{e4} \gamma_l=1-\exp[-ab^{(l-k)}],
\end{eqnarray}
where $a>0,b>1,1\le l\le k$ are parameters of the model. The
parameter $b$ plays the role of the branching number in cascade
models (models of random variables on the branches of hierarchic
models) and $k$ is the maximal number of hierarchy on the tree. An
important difference is that now $b$ is a real number, while in case
of hierarchic trees b should be an integer.  Later we will formulate
the continuous branching version of the model with a single relevant
parameter $V$ defined from the equation
\begin{eqnarray}
\label{e5} b^k=e^V\equiv L\nonumber\\
b\to 1\nonumber\\
L\to\infty
\end{eqnarray}
We will use the notation $L$ in sections II-H,III-B while
investigating the multi-scaling properties of the model.

The model is named as "random walk" model, because, due do Eq.(1),
it is equivalent to the random walks with an amplitude $J^2$, when
the time itself is a random variable see \cite{ba02} and \cite{sa11}
for a simple proof. The random variables $M(t,l)$ are described via
a Markov process, as any time period the transition probability
depends only on the current state. There is a switching according to
Eq.(4), thats why the authors of \cite{cf02} define it as a
"switching" model.

 The distribution of
$M(t,l)$ is chosen to ensure the constraint
\begin{eqnarray}
\label{e6} <M(t,l)>=1,
\end{eqnarray}
where $<>$ means an average.

We can take the log-normal distribution for $M(t,l)$ or normal
distribution for $\epsilon^l_t$, defined as
\begin{eqnarray}
\label{e7} M(t,l)=\exp(\beta \epsilon^l_t),
\end{eqnarray}
where $\beta$ is similar to the inverse temperature in statistical
physics.

We consider a distribution for $\epsilon$:
\begin{eqnarray}
\label{e8} \rho(\epsilon)=\frac{\sqrt{k}}{\sqrt{2\pi
V}}\exp[-\frac{k(\epsilon
-\lambda)^2}{2V}],\nonumber\\
\lambda=-\beta^2/2.
\end{eqnarray}
We have for $|t-t'|<e^V$ the following two-point correlation
function for $u_t$ :
\begin{eqnarray}
\label{e9} <u_tu_{t'}>\sim e^{\beta^2(2V-\ln |t-t'|/\ln b)}.
\end{eqnarray}
The correlation function for $\ln u_t,\ln u_{t'}$ is logarithmic, as
in models discussed in
\cite{bu08},\cite{ca01},\cite{fy09},\cite{ba00}.

The previous expression is derived by observing that $u_t$ and
$u_{t'}$ have identical $M(t,l)$ for $\ln |t-t'|/\ln b $ levels of
hierarchy. The probability that $M(t,l)$ and $M(t',l)$ are identical
is
\begin{eqnarray}
\label{e10}\sim \exp[-|t-t'|e^{-ab^{(l-k)}}].
\end{eqnarray}
Thus $M(t,l)$ and $M(t',l)$ are identical for l-th level of
hierarchy defined through the inequality
\begin{eqnarray}
\label{e11} k-\frac{\ln |t-t'|}{\ln b}<l< k.
\end{eqnarray}
For the rest of the hierarchy levels $M(t,l)$ and $M(t',l)$ are
different.

When
\begin{eqnarray}
\label{e12}\frac{\ln |t-t'|}{V}\ll 1,
\end{eqnarray}
 the majority of the hierarchies have the same $M(t,l)$ and
 $M(t',l)$.

 Alternatively, when
\begin{eqnarray}
\label{e13}1-\frac{\ln |t-t'|}{V}\ll 1,
\end{eqnarray}
 $M(t,l)\ne M(t',l)$ for the the majority of l.

 In section III we derive some more rigorous results.

\section{The statistical physics  of MSM}

\subsection{MSM with general distribution}

Let us consider a general distribution for $\epsilon$
\begin{eqnarray}
\label{e14 } \rho_0(q,\epsilon)=\frac{1}{2\pi
i}\int_{-i\infty}^{i\infty}dh\exp[-h\epsilon+q\phi(h)],
\end{eqnarray}
where $q$ is a parameter indicating some effective length.

We choose the distribution $\rho(\epsilon)$ with some shift
\begin{eqnarray}
\label{e15 } \rho(\epsilon)=\rho_0(\frac{V}{k},\epsilon-\lambda)
\end{eqnarray}
to ensure the constraint given by Eq. (6):
\begin{eqnarray}
\label{e16 } \int_{-\infty}^{\infty} d\epsilon\rho(\epsilon)\equiv
e^{\beta \lambda+\frac{V}{k}\phi(\beta)}=1.
\end{eqnarray}
Thus we take
\begin{eqnarray}
\label{e17 } \lambda=-\frac{V\phi(\beta)}{\beta k}
\end{eqnarray}
where $k$ is our parameter describing the maximal hierarchy level.
Hence, we have the following expression for the correlation
function:
\begin{eqnarray}
\label{e18} <u_tu_{t'}>\sim e^{(V-\ln |t-t'|/\ln
b)\phi(2\beta)+2\phi(\beta)\ln |t-t'|/\ln b }.
\end{eqnarray}

\subsection{The statistical physics versus the dynamics}
Let us define a partition function
\begin{eqnarray}
\label{e19 }
z(i_0,e^V)=\sum_{i=i_0}^{i_0+e^V}x_i\prod_{l=1}^kM(i,l),
\end{eqnarray}
where $M(i,l)$ is chosen from the distribution given by Eq.(15).

Considering Eq.(1) as a dynamic process for a large period of time
M, we define the probability distribution
\begin{eqnarray}
\label{e20 }P(z)=\frac{e^V}{M}\sum_{n=1}^{M/e^V}
\delta_{z,z(1+e^V(n-1),e^Vn)}.
\end{eqnarray}
 We can define a statistical physics as well by
considering $z$ as a partition function for  the 1-d model with
quenched disorder.

The average free energy, denoted as $<\ln Z>$ is:
\begin{eqnarray}
\label{e21 } <\ln Z>\equiv <\ln z(i_0,e^V)>.
\end{eqnarray}
Let us consider a related model with standard distribution for
$\epsilon$ given by $\rho_0(\epsilon)$, without the constraint of
Eq.(4).

We define
\begin{eqnarray}
\label{e22 }
z_0(i_0,e^V)=\sum_{i=i_0}^{i_0+e^V}x_i\prod_{l=1}^kM(i,l),
\end{eqnarray}
where $M_t^l$ are defined through the distribution
$\rho_0(\epsilon)$ and the corresponding free energy is
\begin{eqnarray}
\label{e23 } <\ln Z_0>\equiv <\ln z_0(i_0,e^V)>.
\end{eqnarray}
Eqs.(22) and (23) define a statistical physics model with $e^V$
configurations and special quenched disorder in 1-d space. Later we
will focus on $<\ln Z>$.

 It is easy to check
that
\begin{eqnarray}
\label{e24 } <\ln Z>=<\ln Z_0>-V\phi(\beta). \end{eqnarray}
 Eq.(24) is an exact relation, correct for any value of $\beta$.

 It is easier to solve the model for $<\ln Z_0>$.
 To calculate $<\ln Z_0>$, we will map the model onto the
 REM, and use the standard methods of REM.
One can easily identify the most interesting transition in REM, from
the high temperature phase to the SG phase, by just looking at the
point in the high temperature phase where the entropy disappears.

We will calculate the partition function's moments   $<Z_0^n>$ and
identify them with the $<Z_{REM}^n>$.

\subsection{The moments in MSM model}
First of all we calculate
\begin{eqnarray}
\label{e25 } < (Z_0)^2>=e^Ve^{V\phi(2 \beta)}.
\end{eqnarray}
The cross terms vanishes due to integration by $x_i$.

Let us consider now
\begin{eqnarray}
\label{e26 } <
(Z_0)^{2n}>=\sum_{t_1}..\sum_{t_{2n}}<u_{t_1}..u_{t_{2n}}>.
\end{eqnarray}
While calculating the n-fold sum, we consider two principal
contributions.

The first case is when all $t$ are close to each other and Eq.(12)
is valid. There are $N_1\sim e^V$ such terms and the sum gives
\begin{eqnarray}
\label{e27 } \ln < (Z_0)^{2n}>=V+V\phi(2n\beta)+O(1).
\end{eqnarray}
The second case corresponds to the integration from the regions
where $t_{i+1}-t_i$ are of the same order, and therefore the
condition given by Eq.(13) is satisfied. There are $N_2$ such terms,
\begin{eqnarray}
\label{e28 } \frac{\ln N_2-nV}{V}\ll 1.
\end{eqnarray}
As the vast majority of $M^l$ are different, the average gives
\begin{eqnarray}
\label{e29 } \ln < (Z_0)^{2n}>=nV+nV\phi(2\beta)+O(1).
\end{eqnarray}

\subsection{The corresponding REM}

Consider now $e^V$ energy levels $E_i$  and define the partition
function
\begin{eqnarray}
\label{e30} Z_{REM}=\sum_{i=1}^{e^V}x_ie^{-\beta E_i},
\end{eqnarray}
where $-E_i$ have independent distributions by given Eq. (14) with
$q=V$, and $x_i$ have normal distribution with variance 1.

The moments of the partition function for this model can be
calculated exactly by following \cite{de89} and \cite{sa09}.

These moments  are identical to the expressions given by Eqs. (27)
and (29). We assume that two models with identical integer moments
have an identical free energy as well.

The free energy of the REM model by Eq.(30) can be calculated
rigorously following \cite{sa09}.

At high temperatures, we have the Fisher zeros (FZ) phase with
\begin{eqnarray}
\label{e31} <\ln Z_0>=\frac{1}{2}\ln
(Z_0)^2=V\frac{\phi(2\beta)+1}{2}.
\end{eqnarray}
The transition point is at the point where the entropy disappears:
\begin{eqnarray}
\label{e32} \beta_c\phi'(2\beta_c)=\frac{\phi(2\beta_c)+1}{2}.
\end{eqnarray}
Below this temperature the system is in the SG phase with the free
energy
\begin{eqnarray}
\label{e33} V\beta\phi'(2\beta_c).
\end{eqnarray}
Thus, we have found two phases. The phase given by Eq. (31)
corresponds to the Fisher zero's phase, while that given by Eq.(33)
is the SG phase.

\subsection{Asymmetric distribution}
So far we have considered the case of a symmetric distribution of
$x_i$. Let us now consider the asymmetric case described through the
parameter $\gamma$, where
\begin{eqnarray}
\label{e34} <u_i>=e^{-\gamma V}.
\end{eqnarray}
Now it is possible for the existence of a paramagnetic (PM) phase
with the free energy
\begin{eqnarray}
\label{e35} <\ln Z_0>=\ln <Z_0>=(-\gamma +1+\phi(\beta))V.
\end{eqnarray}
\subsection{Large event}
Let us assume that at the starting moment of time there is a large
event described through the parameter A, while for the other times
Eq.(2) is valid.
 We consider the following partition function
\begin{eqnarray}
\label{e36 }
z_0(i_0,e^V)=-e^{AV}+\sum_{i=i_0+1}^{i_0+e^V}x_i\prod_{l=1}^kM(i,l).
\end{eqnarray}
Now we can have the fourth, ferromagnetic (FM), phase with
\begin{eqnarray}
\label{e37} <\ln |Z_0|>=\ln <Z_0>=AV.
\end{eqnarray}
Actually we can consider an infinite series of time, when after
$e^V$ there is a member $r_i=-e^{AV}$, while for other moments of
time we calculate $r_t$ according to Eqs.(1) and (3).

\subsection{Transition points}
We should choose the proper phase by comparing the expressions given
by Eqs. (31),(33),(35), (37) and then selecting the one which gives
the maximum.

For example, the system transforms from the FZ phase to the PM phase
at
\begin{eqnarray}
\label{e38} e^{(1-\gamma
+\phi(\beta))V}>J^2e^{\frac{V}{2}(1+\phi(2\beta))},
\end{eqnarray}
where $J^2$ is the variance of $x_t$ given by Eq.(2).

\begin{figure}
\centerline{\includegraphics[width=0.95\columnwidth]{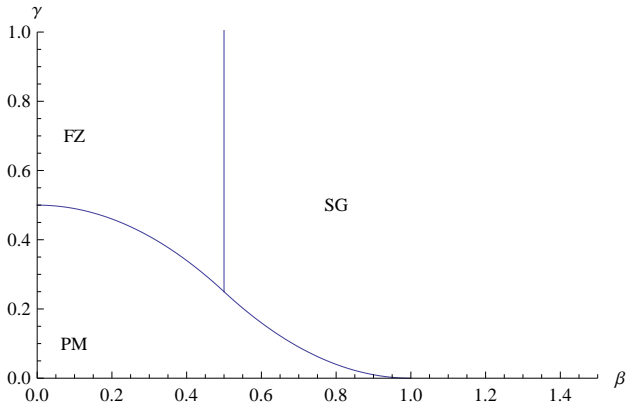}}
\caption{ The phase structure of  the model with asymmetric
distribution of weights.  The case of normal distribution
$\phi(\beta)=\beta^2$. }
  \label{fig1}
\end{figure}

\begin{figure}
\centerline{\includegraphics[width=0.95\columnwidth]{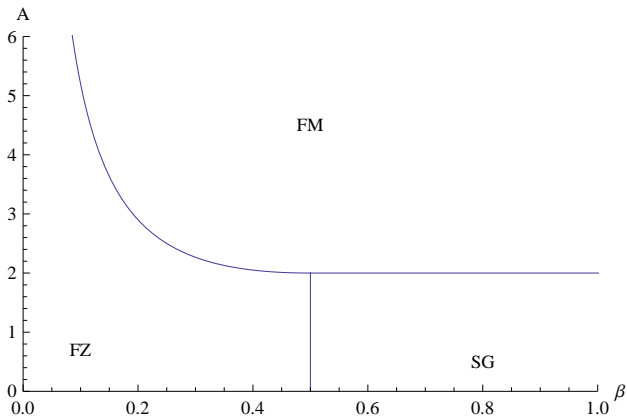}}
\caption{ The phase structure of  the model with a large event.  The
case of normal distribution $\phi(\beta)=\beta^2$.  }
  \label{fig2}
\end{figure}

In Figs. 1,2 we compare the numerics with our analytical results for
the free energy.
\subsection{The scale dependence of the free energy}

Consider again the average distribution of Z, except that, instead
of considering the sum over $e^V$ terms as in Eq.(20), consider the
sum over $l\equiv e^{\alpha V}$ terms,
\begin{eqnarray}
\label{e39 }P(z)=\frac{l}{M}\sum_{n=1}^{M/l} \delta_{z,
z(\alpha,1+l(n-1),ln)},
\end{eqnarray}
where $\epsilon$ have the distribution by Eq.(14).  At high
temperatures we have Fisher zeros (FZ) phase with
\begin{eqnarray}
\label{e40} <\ln Z_0>=V\frac{\phi(2\beta)+\alpha}{2}
\end{eqnarray}
and for the critical point:
\begin{eqnarray}
\label{e41} \beta_c\phi'(2\beta_c)=\frac{\phi(2\beta_c)+\alpha}{2}.
\end{eqnarray}
As $\alpha<1$, the $\beta_c$ decreases with the decrease of
$\alpha$.

\section{The case of continuous branching }
\subsection{The calculation of moments.}
All the formulae in the previous sections have been derived for the
case of general values of b. Consider the case:
\begin{eqnarray}
\label{e42} a=1,k=\frac{V}{\delta v}, b=1+\delta v,\nonumber\\
\delta v\to 0
\end{eqnarray}
and the random variables are distributed according $\rho_0(\delta
v,\epsilon)$.

The the i-th level of hierarchy is unchanged during the period of
time $t$ with a probability:
\begin{eqnarray}
\label{e43} \exp[-te^{v-V}],\nonumber\\
v=i \delta v.
\end{eqnarray}
Multiplying the probabilities in Eq.(43) for different levels of
hierarchy, we obtain
\begin{eqnarray}
\label{e44}
<u_{t_1}u_{t_2}>=\nonumber\\
\prod_{i=1}^k[(1-\exp(-te^v))e^{\delta
v2\phi(\beta)}+\exp(-te^v)e^{\delta v\phi(2\beta)}]=\nonumber\\
\exp[\sum_i\ln((1-\exp(-te^v))e^{\delta
v2\phi(\beta)}+\exp(-te^v)e^{\delta v\phi(2\beta)})]
\end{eqnarray}
Replacing the product by an integral and introducing variables
$x_i=t_i/e^V$, we derive
\begin{eqnarray}
\label{e45}
\int_0^{e^V}dt_1dt_2<u_{t_1}u_{t_2}>=\nonumber\\
e^{V(2(1+\phi(\beta)))}\int_0^1dx_1dx_2\exp[
\int_0^Vdv[\Phi_2(e^v,x_1,x_2)]\nonumber\\
\Phi_2(e^v,x_1,x_2)= e^{-|x_2-x_1|e^{v}}(\phi(2\beta)-2\phi(\beta))
\end{eqnarray}

We get an asymptotic expression with the $e^{-V}$ accuracy  in the
limit $V\to \infty$:
\begin{eqnarray}
\label{e46}
\frac{\int_0^{e^V}dt_1dt_2<u_{t_1}u_{t_2}>}{e^{2V+2V\phi(\beta)}}=\nonumber\\
\int_0^1dx_1dx_2e^{\int_0^{\infty}dv\Phi_2(e^v,x_1,x_2)}.
\end{eqnarray} Similarly, we  derive the expression for the multiple
correlations:
\begin{eqnarray}
\label{e47}
\frac{\int_0^{e^V}dt_1..t_n<u_{t_1}..u_{t_n}>}{e^{nV+n\phi(\beta)}}=\nonumber\\
\int_0^1dx_1..dx_ne^{\int_0^{\infty}dv\Phi_n(e^v,x_1..x_n)}.
\end{eqnarray}The latter expression is $O(1)$, as has been assumed before in Eq.
(29).

 For the 3-point correlation function we obtain
\begin{eqnarray}
\label{e48}
\Phi_3(y,x_1,x_2,x_3)=3+3\phi(\beta)+\nonumber\\
e^{-(x_{12}+x_{23})y}(\phi(3\beta)-3\phi(\beta))\nonumber\\
+e^{-x_{12}y}(1-e^{-x_{23}y})(\phi(2\beta)-2\phi(\beta))\nonumber\\
+(1-e^{-x_{12}y})e^{-x_{23}y}(\phi(2\beta)-2\phi(\beta))]
\end{eqnarray}
where we denote $x_{12}=|x_1-x_2|,x_{23}=|x_2-x_3|$. For n-point
correlation function we need to consider $2^{n-1}$ terms in the
expression of $\Phi_n$.

We can identified this terms with different paths on a tree with
branching number 2, the jumps to the right give a coefficient
$F(x,1)$ and $F(x,-1)$ for the left jump:
\begin{eqnarray}
\label{e49} F(x,1)=\exp[-xe^v], F(x,-1)=(1-\exp[-xe^v])
\end{eqnarray}
 The path is
fractured into clusters, when we have l  subsequent right jumps. We
define
\begin{eqnarray}
\label{e50} f_l=\phi(l\beta)
\end{eqnarray}
We should consider all the paths, the identify the n clusters of the
given path with the length $l_m$ for the m-th cluster. Then we
calculate
\begin{eqnarray}
\label{e51} \Phi_n(e^v,x_1...x_n)=\nonumber\\
\sum_{paths}[\prod_{i=0}^{n-1}F(x_{i,i+1,\alpha})](\sum_{m}f_{l_m}-n\phi(\beta))
\end{eqnarray}

\subsection{Multi-scaling}
If we consider the model for $l=e^{\alpha V}$ period of time and a
normal distribution, we have in the high temperature phase
\begin{eqnarray}
\label{e52} F\equiv \frac{<\ln Z_0>}{\alpha
V}=\frac{1}{2}+\frac{\beta^2}{\alpha}
\end{eqnarray}
and in the SG phase
\begin{eqnarray}
\label{e53} F\equiv \frac{<\ln Z_0>}{\alpha V}=\beta\sqrt
{2/\alpha}.
\end{eqnarray}
The transition point is at
\begin{eqnarray}
\label{e54} \beta_c=\sqrt {\alpha/2}.
\end{eqnarray}
\begin{figure}
\centerline{\includegraphics[width=0.95\columnwidth]{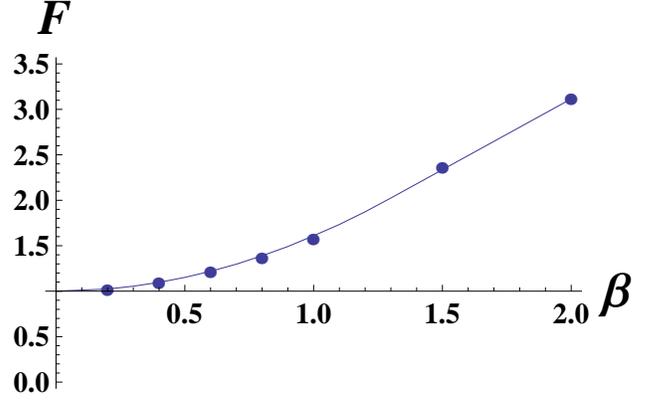}}
\caption{ Free energy F versus an inverse temperature $\beta$ for
the continuous branching model given by Eq.(42) and a normal
distribution. The analytical results by Eqs.(52),(53),
$V=14,N=100000$ are given by smooth lines. }
  \label{fig2}
\end{figure}
In Fig.3 we compare the numerics with our analytical results for the
free energy.

Let us calculate the multi-fractal properties of the MSM. We need to
calculate the moments of the partition function
\begin{eqnarray}
\label{e55}
\int_0^{l}dt_1...dt_n<u_{t_1}...u_{t_n}e^{-n\phi(\beta)V}>\equiv
e^{\ln \frac{l}{L} \xi(n,\beta)}A(L)^nC_n,
\end{eqnarray}
where $\xi(n,\beta)$ defines the multi-scaling, $A(L)$ is some large
numbers while $C_n\sim O(1)$.

While calculating the moments, we slightly modify the formulae of
the previous sections. Instead of Eq.(46) now we obtain
\begin{eqnarray}
\label{e56} \int dt_1..dt_n u(t_1)...u(t_n)=
 \int dx_1..dx_n\times\nonumber\\
 \exp[nV(1+\phi(\beta))+\int_0^{\infty}dv\Phi_n(\frac{l}{L}e^v,x_1...x_n)].
\end{eqnarray}
We consider the case $1\ll l\ll L$. Then, using the equation
\begin{eqnarray}
\label{e57} \Phi_n(0,x_1...x_n)= n\phi(\beta)-\phi(n\beta),
\end{eqnarray}
we derive, integrating by parts:
\begin{eqnarray}
\label{e58} \int_1^{\infty}
\frac{dy}{y}\Phi_n(y\frac{l}{L},x_1...x_n)=\nonumber\\
= (\phi(n\beta)-n\phi(\beta))\ln \frac{l}{L}-\int_0^{\infty}dy\ln
y\Phi_n'(y,x_1...x_n).
\end{eqnarray}
Thus we get a multi-scaling with
\begin{eqnarray}
\label{e59}
 \xi(n,\beta)=n+n\phi(\beta)-\phi(n\beta).
\end{eqnarray}
Considering the moments of
\begin{eqnarray}
\label{e60} z=\sum_{t=1}^{l}u_t,
\end{eqnarray}
where for $u_t$ we use the distribution given by Eq.(15), we obtain
\begin{eqnarray}
\label{e61} \frac{<z^n>}{L^nn!}=e^{\xi(n,\beta)\ln
\frac{l}{L}}\int_0^{1}dx_1...dx_n
 e^{-\int_0^{\infty}dy\ln y\Phi_n'(y,x_1...x_n)}
\end{eqnarray}
and here $x_1...x_n$ are time ordered.
\subsection{The moments for the model with random Boltzman weights.}
Let us calculate now the moments $<z^n>$ for
\begin{eqnarray}
\label{e62} z= \sum_{t=1}^{l}x_tu_t.
\end{eqnarray}
Using Eq.(8) from \cite{sa11}, we derive
\begin{eqnarray}
\label{e63}
\frac{<z^{2n}>}{(LJ)^n}=\frac{2^{n}\Gamma(\frac{1+2n}{2})}{\sqrt{
\pi }}
(\frac{l}{L})^{\xi(n,2\beta)}\nonumber\\
n!\int_0^{0}dx_1...dx_n
 \exp[-\int_0^{\infty}dy\ln y\Phi_n'(y,2\beta,x_1...x_n)],
\end{eqnarray}
where there is a time ordering $t_1<t_2<..t_n$.

\subsection{The multiscaling properties of different phases}

There are no simple order parameters to distinguish the FZ and SG
phases. If we enlarge the free energy expression to the complex
temperatures $\beta=\beta_1+i\beta_2$, then in FZ phase there is a
finite density $\bar \rho$ of partition function zeros, defined
trough the formula \cite{de91a}:
\begin{eqnarray}
\label{e64} \bar \rho(\beta_1,\beta_2)=\frac{1}{2\pi} \frac{<\ln
z(\beta_1,\beta_2)>}{V},
\end{eqnarray}
while in the SG phase this density is zero.

SG and FZ phases have different schemes of replica
symmetry(breaking) \cite{sa01}.

There is a slow relaxation in THE SG phase. Unfortunately there are
no any results about the dynamics of FZ phase to compare.

More interesting is to distinguish two phases  looking the
multiscaling properties. We investigated well the muultiscaling
properties of FZ phase. Let us investigate now the SG phase.

Here there are the results by Gardner and Derrida \cite{de89} about
the the moments of partition functions and the probability
distribution.

Consider again the model with $z=\exp[\alpha V]$ configurations. For
the case
\begin{eqnarray}
\label{e65} \beta> \beta_c\sqrt{\alpha},\beta_c\sqrt{\alpha}>n\beta
\end{eqnarray}
we have, rescaling the result of \cite{de89},
\begin{eqnarray}
\label{e66} \ln <z^n(\alpha)>=n\beta V\sqrt{\alpha}
\end{eqnarray}
In case of the multiscaling the right hand side is proportional to
$\alpha$.

Thus there is a lack of any scaling in the SG phase and we can
distinguish the SG and FZ phases checking the multiscaling property.
We can distinguish FZ and SG phases also looking the tails of the
distributions.

For the FZ phase a simple rescaling of the results of \cite{fy09}
gives for the large $z$:
\begin{eqnarray}
\label{e67} P(z)\sim \frac{1}{z^{1+\frac{\alpha
\beta_c^2}{2\beta^2}}}
\end{eqnarray}
while in the SG phase we used the rescaled result by \cite{de89}
\begin{eqnarray}
\label{e68} P(z)\sim \frac{1}{z^{1+\frac{\sqrt{\alpha}\beta_c
}{\beta}}}
\end{eqnarray}
where $\beta_c/\sqrt{2}$ is the transition point at $\alpha=1$.
Eq.(68) is the result for the REM. For the logarithmic REM a more
accurate expression includes a multiply $\ln{Z}$ on the right hand
side of Eq.(68) \cite{fy12}.

\section{The  dynamic model}

Let us return to the dynamic model given by Eq.(1).
Mapping
\begin{eqnarray}
\label{e68} e^{-\gamma V}=\mu \delta t,\quad J^2=\sigma^2 dt,
\end{eqnarray}
we identify this version of the model at $\beta=0$ as a finite time
version of driven Brownian motion \cite{de02}. In case of simple
random walks with $\beta=0$, there is a single phase.  In the model
given by Eq.(1) with $\beta\ne 0$ we have an intrinsic large
parameter $L\equiv e^V$, describing the effective number of
configurations in the model. In this way the statistical physics
enters into the dynamical problem. Contrary to the driven Brownian
motion and the
 Heston model  \cite{he}, we have 4 different phases in the MSM model.
 The same is the situation with other models of multi-scaling
 random walks \cite{ba00}.

 To identify the choice amongst the two phases (FZ
versus PM) we consider
\begin{eqnarray}
\label{e69} C=\frac{\sigma}{|\mu|\sqrt{T}}.
\end{eqnarray}
 When $C\ll 1$, the system is in the PM phase. Otherwise at $C\gg 1$ we have the FZ phase.

\section{Conclusion}

We considered the dynamic Markov switching multi-fractal  models
(MSM) and connected them with a new class of solvable statistical
physics models of quenched disorder in one dimension. In these
models there is both translational invariance and general
distribution of disorder. While the multiscaling in FZ phase has
been investigated before, nor the statistical physics properties
properties, neither the phase structure has been investigated
before. We found the exact phase structure of the model. At
different phases there should be different distributions $P(z)$. In
case of a symmetric random walk there are two phases in the
considered model. At small parameters $\beta$, the model is in the
phase with non-zero density of Fisher's zeros. At high $\beta$, the
system is in the spin-glass phase, a pathologic phase with a slow
relaxation dynamics. For an asymmetric distribution of $x_i$ there
is a possibility for the third, paramagnetic, phase. Slight
modification of the model allows the existence of the fourth,
ferromagnetic, phase.
It is possible to distinguish different phases measuring the
multi-scaling properties of the model. The multi-scaling is broken
in the SG phase.
 We also introduced a continuous hierarchy branching
version of the MSM, gave expressions for the moments of the
partition function, and calculated the multi-scaling indices.

For applications it is important to calculate the fractional moments
of the partition function. Perhaps we can use expressions for
integer moments and use some approximate methods of extrapolation.
Another interesting open problem is to investigate the dynamics of
the model, looking for a new phase transition point in the dynamics,
as is the case of the spherical spin-glass model \cite{cr93}.

 DBS thanks  Academia Sinica for financial
support, Y. Fyodorov, S. Jain, Th. Lux, D. Sornette and O. Rozanova
for discussions.

\end{document}